# Structural, magnetic, and electronic evolution of the spin-ladder system BaFe$_2$S$_{3-x}$Se$_x$ with isoelectronic substitution

Jia Yu[1], Meng Wang[1*], Benjamin A. Frandsen[2,3#], Hualei Sun[1], Junjie Yin[1], Zengjia Liu[1], Shan Wu[4,2], Ming Yi[4,5], Zhijun Xu[6], Arani Acharya[4], Qingzhen Huang[6], Edith Bourret-Courchesne[2], Jeffrey W. Lynn[6], and Robert J. Birgeneau[2,4]

[1]*School of Physics, Sun Yat-sen University, Guangzhou, Guangdong 510275, China*
[2]*Materials Sciences Division, Lawrence Berkeley National Laboratory, Berkeley, California 94720, USA*
[3]*Department of Physics and Astronomy, Brigham Young University, Provo, Utah 84602, USA*
[4]*Department of Physics, University of California, Berkeley, California 94720, USA*
[5]*Department of Physics and Astronomy, Rice University, Houston, TX 77005, USA*
[6]*NIST Center for Neutron Research, National Institute of Standards and Technology, Gaithersburg, Maryland 20899, USA*

We report experimental studies of a series of BaFe$_2$S$_{3-x}$Se$_x$ ($0 \leq x \leq 3$) single crystals and powder specimens using x-ray diffraction, neutron diffraction, muon spin relaxation, and electrical transport measurements. A structural transformation from *Cmcm* (BaFe$_2$S$_3$) to *Pnma* (BaFe$_2$Se$_3$) was identified around $x = 0.7 \sim 1$. Neutron diffraction measurements on the samples with $x = 0.2, 0.4$, and $0.7$ reveal that the Néel temperature of the stripe antiferromagnetic order is gradually suppressed from ~120 to 85 K, while the magnitude of the ordered Fe$^{2+}$ moments shows very little variation. Similarly, the block antiferromagnetic order in BaFe$_2$Se$_3$ remains robust for $1.5 \leq x \leq 3$ with negligible variation in the ordered moment and a slight decrease of the Néel temperature from 250 K ($x=3$) to 225 K ($x=1.5$). The sample with $x=1$ near the *Cmcm* and *Pnma* border shows coexisting, two-dimensional, short-range stripe- and block-type antiferromagnetic correlations. The system remains insulating for all $x$, but the thermal activation gap shows an abrupt increase when traversing the boundary from the *Cmcm* stripe phase to the *Pnma* block phase. The results demonstrate that the crystal structure, magnetic order, and electronic properties are strongly coupled in the BaFe$_2$S$_{3-x}$Se$_x$ system.

## I. Introduction

High-temperature (HT$_c$) superconductors, i.e. the cuprates, the iron pnictides, and the iron chalcogenides, have stimulated great interest in the exploration of new superconducting materials[1-3]. The interplay among the structure, magnetism, and electronic state is crucial for understanding HT$_c$ superconductivity. Most HT$_c$ superconductors exhibit a quasi-two-dimensional structure with layers of CuO$_2$, FeAs, or FeS/Se playing an essential role for the superconductivity[2-4]. Antiferromagnetic (AF) spin fluctuations are also believed to be of great importance for the superconducting paring mechanism in both copper-based and iron-based HT$_c$ systems [5-7]. One major difference between the parent compounds of the two families is that the cuprates are Mott insulators, while the iron pnictides/chalcogenides show semimetallic behaviors[2,8,9]. Many of the insulating phases that have been found in the iron-based family do not appear to be closely connected to the superconducting properties [10-15].

The recent discoveries of pressure-induced superconductivity in BaFe$_2$S$_3$ and BaFe$_2$Se$_3$ have somewhat altered this understanding of the dichotomy between the Mott-insulating cuprates and semimetallic iron-based HT$_c$s. BaFe$_2$S$_3$ and BaFe$_2$Se$_3$ are both Mott insulators at ambient conditions, but under hydrostatic pressures of ~10 GPa, the insulating phase is replaced by a metallic phase and superconductivity appears. Therefore, these 123-type iron chalcogenides are considered to be one of the iron based HT$_c$ superconducting systems [27,28], and yet they share a common Mott insulating starting point with the cuprates. These materials are therefore of significant interest to the HT$_c$ community.

BaFe$_2$S$_3$ and BaFe$_2$Se$_3$ belong to the broader class of 123-type $A$Fe$_2$(S/Se)$_3$ ($A$ = K, Rb, Cs, or Ba) materials [16-19]. In contrast to the widely studied layered iron-based superconductors, the $A$Fe$_2$(S/Se)$_3$ system exhibits a quasi-one-dimensional (Q1D) ladder structure comprised of Fe atoms. The crystal structure of most of the $A$Fe$_2$(S/Se)$_3$ compounds belongs to the orthorhombic space group *Cmcm*, and the Fe ladders have a stripe-type AF ground state. However, the structure of BaFe$_2$Se$_3$ has a slightly lower symmetry that belongs to the orthorhombic space group *Pnma*, resulting in a slight out-of-plane tilt of the Fe ladders. The distorted Fe ladders exhibit a block-type AF ground state [Fig. 1(a)]. Great efforts have been made to identify superconducting phases in this Q1D system[20-26], including experimental and theoretical research under both ambient conditions and pressure[29-42]. To date, superconductivity has only been found in BaFe$_2$S$_3$ and BaFe$_2$Se$_3$ under pressure around 10 GPa.

In BaFe$_2$S$_3$, the *Cmcm* structure remains unchanged with pressure up to at least ~17 GPa [43]. In contrast, the *Pnma* structure of BaFe$_2$Se$_3$ transforms to *Cmcm* around 3.5-4.4 GPa [44], after which superconductivity emerges at higher pressures. Hence, the *Cmcm* structure may be viewed as the parent structure of the superconductivity in the 123 system. However, there is no sign of transformation from the block AF order to the stripe AF order up to ~6.8 GPa in BaFe$_2$Se$_3$[45]. The nature of the magnetic ground state in the BaFe$_2$Se$_3$ phase proximate to the SC state is not clear. Reasonable possibilities include an unchanged block AF state, competing and coexisting block and stripe AF states, or a complete transition to stripe AF order at pressures beyond what has previously been explored.

One way to investigate how the two phases connect is to explore the solid solution of S and Se in BaFe$_2$(S,Se)$_3$, effectively tuning the nuclear and magnetic structure from the *Pnma* and block AF order in BaFe$_2$Se$_3$ to the *Cmcm* and stripe AF order in BeFe$_2$S$_3$. This isovalent substitution can be regarded as a type of chemical pressure. A recent study has established a chemical substitution $x$ versus temperature $T$ phase diagram from magnetic susceptibility measurements[46]. However, a comprehensive investigation of the structural, magnetic, and electronic evolution across the $x-T$ phase diagram of BaFe$_2$S$_{3-x}$Se$_x$ is still lacking, representing a significant gap in our fundamental knowledge of this Q1D system.

In this paper, we report comprehensive studies on the evolution of the structural, magnetic, and electronic properties through the application of chemical pressure via isovalent substitution of Se for S in BaFe$_2$S$_{3-x}$Se$_x$. Powder x-ray diffraction (PXRD) measurements show that the structural transition from *Cmcm* to *Pnma* occurs between $x = 0.7 \sim 1.0$. Neutron powder diffraction (NPD) experiments reveal that the S-rich samples ($x \leq 0.7$) maintain the stripe AF order in the *Cmcm* phase with little change in the ordered moment and only a slight decrease in $T_N$. Similarly, the Se-rich samples ($x \geq 1.5$) preserve the block AF order with negligible change in the ordered moment and only a slight decrease in $T_N$ as Se content decreases from $x = 3$ to 1.5. A sample of BaFe$_2$S$_2$Se, which is in the *Pnma* phase but is very close to the border with the *Cmcm* phase, shows short-range, two-dimensional (2D) correlations of both the block type and stripe type. Muon spin relaxation (μSR) experiments manifest that static magnetism develops throughout the full volume for samples across the entire substitution series, even in those samples showing only short-range order. Resistivity measurements reveal an abrupt enhancement of the thermal activation gap with the appearance of the *Pnma* structure and block AF order for $x \geq 1$. Using these combined results, we present a complete phase diagram depicting the evolution of the crystal structure, magnetic order, and electrical transport properties, revealing a strong coupling among these various degrees of freedom.

## II. Experiments

Single crystals of BaFe$_2$S$_{3-x}$Se$_x$ ($x$ = 0, 0.2, 0.4, 0.5, 0.7, 1.0, 1.3, 1.5, 1.7, 2.5, 3.0) were grown from self-flux using the Bridgman method[14]. In addition, powder samples were synthesized for $x$ = 1.5, 2, 2.33, 2.67, and 3.0 following the procedure in ref. [25]. Room temperature PXRD experiments were performed using a homemade diffractometer with a two-dimensional detector. Neutron powder diffraction measurements of ground up single crystals with $x = 0.2{\sim}0.7$ were conducted on the BT1 at the NIST Center for Neutron Research (NCNR) using a monochromatic beam with a wavelength of $\lambda = 2.0775$ Å. The sample with $x = 1$ was measured on the NOMAD time-of-flight powder diffractometer at the Spallation Neutron Source at Oak Ridge National Laboratory (ORNL). The powder samples with $1.5 \leq x \leq 3$ were studied with neutron diffraction on the HB-2A instrument at the High Flux Isotope Reactor at ORNL using a monochromatic beam with a wavelength of $\lambda = 2.4104$ Å. Elastic neutron scattering measurements of single crystalline BaFe$_2$S$_2$Se were carried out on the BT-7 thermal triple-axis spectrometer at NCNR, NIST[47]. Uncertainties were indicated are statistical in origin and represent one standard deviation. Muon spin relaxation measurements were performed on the M20 beamline at the Centre for Molecular and Materials Science at TRIUMF in Vancouver, Canada. Electrical transport properties of the single crystal specimens were measured on a Quantum Design Physical Property Measurement System (PPMS) with a four-probe method to eliminate contact resistance.

## III. Results

### A. Powder x-ray diffraction measurements

PXRD at room temperature was employed to examine the sample homogeneity and detect the structural transition in BaFe$_2$S$_{3-x}$Se$_x$. We plot the PXRD patterns for all compositions in Fig. 1(b). The calculated peak positions for the BaFe$_2$S$_3$ structure are marked at the bottom and a zoomed-in plot for the dashed box area is attached to the right. The observed peak positions of BaFe$_2$S$_3$ agree well with the calculation. Upon introduction of the larger Se atoms, the peaks shift to lower angles due to the expanded lattice. The continuous peak shifting from BaFe$_2$S$_3$ to BaFe$_2$Se$_3$ indicates an effective substitution. The red arrows in the zoomed-in plot show the evolution of the (3, 1, 0) peak of the *Cmcm* structure, which is forbidden in the *Pnma* structure for BaFe$_2$Se$_3$. This peak vanishes for $x$ = 1.0, demonstrating that the structural transition occurs between $x = 0.7$ and $x = 1.0$.

The PXRD patterns of the samples with $0 \leq x \leq 0.7$ and those with $1 \leq x \leq 3$ are fit with the space groups *Cmcm* and *Pnma*, respectively. The lattice parameters normalized to the values for BaFe$_2$S$_3$ are presented in Fig. 1(c), where $a_0$ = 8.795(6) Å, $b_0$ = 11.244(7) Å, and $c_0$ = 5.292(3) Å are the lattice constants of BaFe$_2$S$_3$. There is a transformation of coordinates between the *Cmcm* and *Pnma* structures, so for the sake of consistency, we designate the $a$, $b$, $c$ lattice parameters in Fig. 1(c) to be along the ladder rung, perpendicular to the ladder plane, and along the ladder direction, respectively. The lattice parameters increase continuously from BaFe$_2$S$_3$ to BaFe$_2$Se$_3$, except for BaFe$_2$S$_{2.3}$Se$_{0.7}$. The contraction of the lattice constants for $x$ = 0.7 has been confirmed on a high-resolution x-ray diffractometer (PANanalytical Empyrean) and by the neutron powder diffraction measurements to be described subsequently. The anomaly at $x$ = 0.7 is possibly associated with a release of the residual stress that accompanies the first order structural transition. We note that the lattice parameter perpendicular to the ladder plane ($b$ in Fig. 1(c)) changes much faster than the other two in pressurized BaFe$_2$S$_3$ and BaFe$_2$Se$_3$ and Ni-doped BaFe$_2$Se$_3$[27, 29,48]. However, the present results for chemical pressure show that the change in lattice parameter along the rung direction is very similar to the change perpendicular to the ladder plane. This suggests that the application of chemical pressure may introduce additional residual stress into the structure.

## B. Neutron scattering measurements

TABLE 1. The refined structural parameters for BaFe$_2$S$_{3-x}$Se$_x$ ($x$ = 0.2, 0.4, and 0.7) from the NPD data at 6 K. The agreement factors for the compounds with $x$ = 0.4 and 0.7 are $R_P$ = 13.9%, $wR_P$ = 14.2%, $\chi^2$ = 4.08%, and $R_P$ = 16.3%, $wR_P$ = 16.1%, $\chi^2$ = 1.51%, respectively.

| | | | $x$ = 0.2 | | | | $x$ = 0.4 | $x$ = 0.7 |
|---|---|---|---|---|---|---|---|---|
| | | $R_P$=11.7% | $wR_P$=12.7% | | $\chi^2$=2.08% | | $a$=8.7950(4) Å | $a$=8.8153(5) Å |
| | | $a$=8.7596(2) Å | $b$=11.2073(2) Å | | $c$=5.2756(1) Å | | $b$=11.2491(6) Å | $b$=11.2666(6) Å |
| | | | | | | | $c$=5.2859(2) Å | $c$=5.2888(3) Å |
| | Ordered moment ($\mu_B$) | | | | 1.30(3) | | 1.40(4) | 1.34(4) |
| Atom | Site | x | y | z | Occ. | | Occ. | Occ. |
| Ba | 4c | 0.5000 | 0.1893(3) | 0.2500 | 1.0000 | | 1.0000 | 1.0000 |
| Fe | 8e | 0.3466(2) | 0.5000 | 0.0000 | 0.994(7) | | 1.002(8) | 0.986(8) |
| S1 | 4c | 0.5000 | 0.6139(6) | 0.2500 | 0.910(3) | | 0.889(4) | 0.679(5) |
| Se1 | 4c | 0.5000 | 0.6139(6) | 0.2500 | 0.058(3) | | 0.094(4) | 0.173(5) |
| S2 | 8g | 0.2035(4) | 0.3766(4) | 0.2500 | 0.927(7) | | 0.818(9) | 0.79(1) |
| Se2 | 8g | 0.2035(4) | 0.3766(4) | 0.2500 | 0.073(7) | | 0.182(9) | 0.21(1) |

To gain a better understanding of the detailed evolution of the structure and magnetism, high-quality NPD data were collected for samples spanning the full composition series. Representative NPD patterns are displayed in Fig. 2. For the S-rich compounds of $x$ = 0.2, 0.4, and 0.7, the nuclear peaks can be well described by the *Cmcm* structure. The patterns at 6 K show clear additional peaks attributable to AF order in Figs. 2(a)-2(c). These magnetic peaks can be indexed with a propagation vector $k$ = (0.5, 0.5, 0) and quantitatively fit with a model of the stripe AF order in BaFe$_2$S$_3$. The ordered moments are refined to be 1.30±0.03, 1.43±0.04, and 1.34±0.04$\mu_B$ per Fe, respectively. The slightly enhanced ordered moments from 1.29±0.03$\mu_B$ in BaFe$_2$S$_3$ to 1.43±0.04$\mu_B$ in BaFe$_2$S$_{2.6}$Se$_{0.4}$ may be due to the enlarged lattice constants and enhanced electronic localization.

The refined parameters for the S-rich compounds are summarized in Table 1. The compositions are approximately BaFe$_2$S$_{2.76}$Se$_{0.20}$, BaFe$_2$S$_{2.53}$Se$_{0.46}$, and BaFe$_2$S$_{2.26}$Se$_{0.59}$, close to the nominal compositions. Interestingly, the Se content on the S2 site is greater than that on the S1 site for each compound, indicating preferential occupation of Se on the S2 site. The S1 site is a key component for stabilizing the Fe ladders, while the S2 site helps connect ladders within a given plane. The greater Se content on the S2 site is likely responsible for the faster expansion of the lattice along the rung direction and perpendicular to the ladder plane. The atom occupancies in BaFe$_2$S$_{2.3}$Se$_{0.7}$ show about 15% vacancies on the S1 site. Thus, the release of stress is achieved by introducing vacancies on the S1 site. A similar effect of vacancy-enabled stress release is observed in ZnO thin films[49].

Diffraction patterns for several Se-rich compounds are displayed in Figs. 2(d)-2(f). Small impurity peaks corresponding to FeSe are observed, particularly around $2\theta$ = 26°. Of note is the appearance of additional Bragg peaks at low temperature, as shown for BaFe$_2$S$_{0.33}$Se$_{2.67}$, signaling the development of long-range AF order. A particularly strong peak at $2\theta = 17°$ corresponding to $Q = (0.5, 0.5, 0.5)$ in the reciprocal lattice units of the *Pnma* structure is observed, consistent with the block-type AF order in BaFe$_2$Se$_3$. For the compositions of $x = 1.5$ and 2.67 showing long-range AF order, the ordered moment refines to be $2.8 \pm 0.1\mu_B$, with little systematic variation as a function of composition.

Interestingly, the powder sample of BaFe$_2$S$_{0.67}$Se$_{2.33}$ ($x = 2.33$) shows no long-range magnetic order, even though compositions on either side of it in the substitution series do. Nevertheless, short-range correlations are clearly present at low temperature, as indicated by the diffuse feature centered around $2\theta = 17°$ seen in Fig. 2(e). More detailed analysis to be presented subsequently reveals that this diffuse scattering results from a combination of 2D and 3D short-range block AF correlations. The origin of the short-range order in this particular sample is not completely understood, but we propose that Fe vacancies may be at least partially responsible. The refined Fe occupancy for $x = 2.33$ is $98 \pm 0.5\%$, while the Fe occupancies for all other Se-rich samples showing long-range AF order refine robustly to 100%. In the case of BaFe$_2$S$_{1.5}$Se$_{1.5}$, we synthesized two samples, one which had long-range block AF order and a full Fe occupancy, and the other which had only short-range block correlations and $96 \pm 0.5\%$ Fe occupancy. The sample of BaFe$_2$S$_2$Se, which we will discuss in greater detail subsequently, likewise exhibits only short-range AF correlations and also has a refined Fe occupancy of $96 \pm 0.5\%$. The common trend of Fe deficiency in samples lacking long-range AF order suggests that Fe vacancies and other microscopic, sample-dependent details can destroy the long-range block AF order in the *Pnma* structure. The delicate nature of the long-range order is likely related to the low dimensionality of the system. Nevertheless, the underlying AF interactions remain strong, resulting in robust short-range correlations even in the absence of long-range order.

To extract the evolution of $T_N$ as a function of Se doping, magnetic Bragg peak intensities are shown as a function of temperature for various compositions in Fig. 3. For the S-rich compounds, the stripe-type AF transition temperature is suppressed from 120 K in BaFe$_2$S$_3$ to 107 K in BaFe$_2$S$_{2.8}$Se$_{0.2}$, 101 K in BaFe$_2$S$_{2.6}$Se$_{0.4}$, and 85 K in BaFe$_2$S$_{2.3}$Se$_{0.7}$, as shown in Fig. 3(a), despite the slight enhancement of the ordered moment[27,28]. Figure 3(b) displays the representative peak intensities at $Q = (0.5, 0.5, 0.5)$ for BaFe$_2$S$_{0.33}$Se$_{2.67}$ and BaFe$_2$S$_{1.5}$Se$_{1.5}$, showing clear transitions around 250 K and 225 K, respectively, revealing a slight suppression of $T_N$ with substitution away from the Se end member. For BaFe$_2$S$_{0.67}$Se$_{2.33}$ with short-range AF order, the diffuse intensity gradually develops as the temperature is lowered, also seen in Fig. 3(b).

The magnetic behavior close to the structural transition holds crucial information for understanding how the magnetic structure evolves from stripe-type to block-type with Se substitution. Therefore, we conducted elastic neutron scattering studies on a single crystal of BaFe$_2$S$_2$Se. The indexing for BaFe$_2$S$_2$Se is based on the space group *Pnma* due to the structural transformation. A broad magnetic peak centered at $Q = (0.5, 1, 0.5)$ is seen for the data collected at and below 40 K [Fig. 4(a)]. Similar broad peaks appear at $Q = (1.5, 1, 0.5)$ and $(2.5, 1, 0.5)$, as presented in Figs. 4(b) and 4(c). All of these peaks are consistent with the stripe AF order observed in BaFe$_2$S$_3$, indicating that short-range stripe-type correlations persist into the *Pnma* structural phase. To estimate the correlation length of the short-range magnetic order, we fit Gaussian functions to the magnetic peaks for $Q = (0.5, 1, 0.5)$, $(1.5, 1, 0.5)$, and $(2.5, 1, 0.5)$ at 2.5 K, shown as the black solid curves in Figs. 4(a)-3(c). The cyan star symbols in Fig. 4(b) represent the nuclear peak for $Q = (3, 2, 1)$ measured without pyrolytic graphite (PG) filter. The width of the nuclear peak can be taken as the resolution of the instrument. From the Gaussian fits, the short-range magnetic correlation lengths within the ladder planes are estimated to be $\xi_{1.5} = 13 \pm 2$ Å for $H = 1.5$, $\xi_{0.5} = 20 \pm 2$ Å for $H = 0.5$, and $\xi_{2.5} = 29 \pm 4$ Å for $H = 2.5$[15]. The effects of the instrument resolution for $H = 0.5$ and 2.5 are not significant and have been neglected.

To check the magnetic order along the direction perpendicular to the ladder plane, scans along $Q = (H, 1, 0.5)$ with $H$ from 0 to 3 at 2.5 and 113 K were conducted. The difference of the scattered intensity between 2.5 and 113 K, which includes the magnetic scattering, is plotted in Fig. 4(d). Aside from a slight monotonic decrease in scattered intensity with $H$, no modulation of the magnetic peak along the $H$ direction is observed, indicating a complete lack of ordering perpendicular to the ladder plane. Therefore, we conclude that the short-range stripe-type magnetic order in BaFe$_2$S$_2$Se is 2D. The scans mapped in the $(H, 2K, K)$ plane in Fig. 4(e) reveal a ridge of intensity along the (H, 1, 0.5) direction, further confirming the absence of the magnetic correlations along the $a$ axis (out-of-ladder plane direction). Figure 4(f) shows the temperature dependence of the intensities of the magnetic peaks at $Q = (1.5, 1, 0.5)$ and $(2.5, 1, 0.5)$ (left vertical axis) and the full widths at half maximum of the peak at $Q = (1.5, 1, 0.5)$ (right vertical axis). From the intensities, we determine the crossover temperature to short-range stripe correlations in BaFe$_2$S$_2$Se to be $\sim$49 K. The broadened peak widths at the higher temperatures suggest that thermal fluctuations weaken the short-range magnetic correlation and shorten the correlation lengths.

The scattering plane selected for the single-crystal neutron diffraction measurements of BaFe$_2$S$_2$Se did not provide access to the block AF positions in reciprocal space, precluding any conclusions about block AF order in this sample. However, we performed additional neutron powder diffraction measurements on ground up single crystals from the same batch of BaFe$_2$S$_2$Se that probe a more complete volume of reciprocal space. We performed fits to the diffuse magnetic scattering using the diffpy.magpdf package [51]. Corrections were made to minimize artifacts from thermal shifts of the Bragg peaks, although these artifacts could not be completely removed. The results are shown in Fig. 5(a), which has also been shown in our recent review paper [52]. The strong feature centered around 0.7 Å$^{-1}$, which is approximately the $Q = (0.5, 0.5, 0.5)$ position, corresponds to short-range block-type correlations, confirming the presence of both stripe and block correlations in this compound. The asymmetric peak shape arises from the 2D nature of the correlations. The weaker feature around 1.25 Å$^{-1}$ is due to the stripe-type correlations described in the previous discussion. The weak features at higher $Q$ correspond to the block phase. The 2D correlation lengths for the stripe and block correlations determined from the fits are $\sim$1 and $\sim$3 nm, respectively. The 2D stripe-type correlation length obtained from the powder data is somewhat smaller than that obtained from the single-crystal data, but they are qualitatively consistent. Based on the diffraction data, the coexistence of these two types of magnetic correlations does not seem to originate from phase separation in the sample between S-rich regions with the *Cmcm* structure showing stripe correlations and Se-rich regions with the *Pnma* structure showing block correlations. Instead, the sample appears to be structurally and compositionally homogeneous, revealing an interesting situation in which compounds in the *Pnma* structural phase that are near the boundary with the *Cmcm* phase host coexisting stripe- and block-type interactions.

For comparison, the diffuse magnetic scattering for BaFe$_2$S$_{0.67}$Se$_{2.33}$ is shown in Fig. 5(b). The main peak from the block-type correlations, clearly visible around 0.7 Å$^{-1}$, is significantly sharper than the corresponding peak for BaFe$_2$S$_2$Se. We found that a purely 2D model could not accurately describe this peak or the rest of the diffuse scattering pattern; instead, a model consisting of both 2D and 3D short-range correlations was necessary to obtain a good fit. From the fit, we estimate the in-plane correlation length to be $11 \pm 3$ nm and the out-of-plane correlation length to be $6 \pm 2$ nm. No diffuse intensity is observed at the stripe-type wave vector around 1.25 Å$^{-1}$, indicating that no observable stripe-type correlations are present in this compound.

### C. Muon spin relaxation measurements

As a probe of local magnetic fields, μSR is a valuable complement to neutron scattering methods. In particular, its sensitivity to the volume fraction of magnetically ordered phases provides important new information not easily obtained by NPD. µSR measurements of BaFe$_2$S$_3$ and BaFe$_2$Se$_3$ under pressure have been reported previously[45,50]. Here, we focus on measurements across the S-Se phase diagram at ambient pressure. Representative zero-field (ZF) µSR asymmetry spectra for BaFe$_2$Se$_3$ are displayed in Fig. 6 (a). Above $T_N$, the asymmetry exhibits a gentle, exponential-type relaxation with time. As the temperature is lowered toward the transition, the relaxation rate increases, indicating a critical slowing down of spin dynamics as the transition is approached. As the transition is traversed starting around 250 K, muons landing in regions of the sample with static magnetic order experience a very rapid spin depolarization, resulting in an apparent loss of asymmetry at $t = 0$. Once magnetic order extends through the full volume of the sample, only the "1/3" asymmetry tail remains. Low-energy excitations out of the ordered state cause the tail to relax, as seen at 200 K. As the temperature is lowered, these excitations are frozen out and the asymmetry remains constant in time, as seen at 2 K.

The solid curves in Fig. 6(a) represent fits using a model consisting of two exponential functions with temperature-dependent amplitudes and relaxation rates, corresponding to the paramagnetic volume fraction and the ordered volume fraction of the sample, respectively. The ordered volume fraction can be extracted from these fits. The temperature dependence of the ordered volume fraction is displayed for representative compounds in Figs. 6(b-e). In all cases, including compositions not shown in Fig. 6, 90% or more of the muons experience static magnetism at base temperature. These results were confirmed quantitatively by calibration measurements performed in a weak transverse field (data not shown). The remaining ~10% of muons that land in paramagnetic regions for some of the samples may be due to muons stopping in the sample holder or possibly some small nonmagnetic inclusions in the samples. In any case, the µSR measurements confirm that static magnetism extends through the bulk of all samples studied. We note that samples with intermediate compositions tended to have somewhat broader transitions in temperature compared to the end member compounds, perhaps suggesting a slight distribution of $T_N$ due to microscopic variations in chemical composition.

Interestingly, the µSR results demonstrate that the short-range order in BaFe$_2$S$_2$Se and BaFe$_2$S$_{0.67}$Se$_{2.33}$ also extends through the full volume fraction. In particular, the results for BaFe$_2$S$_2$Se shown in Fig. 6(c) indicate that static magnetic correlations begin to develop below 100 K. From the results of neutron diffraction on single crystals, we know the stripe correlations first being to develop below 49 K; therefore, we attribute the correlations between 49 K and 100 K to the block phase. Significantly, the magnetic volume fraction has nearly attained its maximum value at 50 K before the stripe correlations develop, demonstrating that the block correlations are not confined just to a partial volume fraction of the sample. Instead, the block correlations appear to occupy nearly the full sample volume even before the stripe correlations develop, allowing us to conclude that there is no significant macroscopic segregation of the sample into block-type correlations (which would presumably be rich in Se) and stripe-type correlations (which would be rich in S). Instead, the sample is homogeneous within the sensitivity of µSR. This supports the structural homogeneity indicated by the XRD and NPD analysis and points to a genuine coexistence of the stripe- and block-type correlations.

### D. Electrical transport measurements

To complement our investigations of the structural and magnetic evolution across the S-Se composition series, we performed electrical transport measurements on a series of samples to probe the evolution of the electronic state. This builds upon previous experiments showing that the thermal activation gap of the insulating state for BaFe$_2$Se$_3$ is much larger than that for BaFe$_2$S$_3$[21,53]. The resistivity versus temperature ($\rho - T$) curves for various compositions $x$ are illustrated in Fig. 7(a). Although all the compounds show insulating behavior, they can be classified into two groups with quantitatively different behavior: $x \leq 1.0$ and $x > 1.0$. The $\rho$-$T$ curve for each sample can be fit using the activation-energy model $\rho(T) = \rho_0 exp(E_a/k_B T)$ as shown in Fig. 7(b), where $\rho_0$ is a prefactor, $E_a$ is the thermal activation gap, and $k_B$ is the Boltzmann's constant.

The activation energy gap obtained from fits over the temperature range $100\,\text{K} < T < 200\,\text{K}$ are shown as the red diamonds in Fig. 8(a). A smaller energy gap of $40 \sim 50$ meV for $x \leq 1$ is clearly distinct from a larger gap of $100 \sim 120$ meV for $x > 1$. To evaluate any possible influence of the AF order on the electrical transport properties, we performed equivalent fits over the temperature range $40 \sim 100$ K (mostly below $T_N$) for $x \leq 1.0$. The results, shown as the black squares in Fig. 8(a), reveal only small differences between the transport properties in the paramagnetic state and those in the AF state. For $x > 1.0$, we did not attempt to fit the activation energy model to the $\rho - T$ curves over the lower temperature range, since the large resistivity prevented the collection of data at temperatures much below 100 K. We observe that the activation energy gap persists in the range of $100 \sim 120$ meV for $1.5 \leq x \leq 3.0$, approximately $2 \sim 3$ times the energy gap for $x \leq 1.0$. Considering this fact together with the minimal change across the AF transition, we conclude that the structural transition plays a dominant role in determining the electrical transport properties in the isovalent substitution process.

## IV. Discussion and summary

From the combined results of these experimental studies, we have assembled in Fig. 8 a comprehensive phase diagram showing the evolution of the crystal structure, magnetic order, and thermal activation gap as a function of Se content $x$. Part of this phase diagram was included in our review paper[52]. The crystal structure transitions from *Cmcm* to *Pnma* between $x = 0.7$ and $1.0$. Concomitantly, the activation gap is enhanced abruptly beginning with $x = 1.0$, consistent with the nature of a first-order structural transition. The Néel temperatures of both the stripe and block AF orders are suppressed slightly via chemical substitution away from the $BaFe_2S_3$ and $BaFe_2Se_3$ end members, consistent with the behavior of a bicritical point suggested from a recent magnetic susceptibility study[47]. The influence of the structural transition on the ordering type and temperature is similar to that seen in the 2D iron pnictide $Ba(Fe_{1-x}Mn_x)_2As_2$[54]. We also observe short-range, 2D magnetic order of both the stripe and block type for $x = 1.0$, suggesting an intermediate state between the long-range stripe and block AF orders in $BaFe_2S_3$ and $BaFe_2Se_3$, respectively. Together with recently reported results on thermal conductance in $BaFe_2S_{3-x}Se_x$, this suggests that the magnons from stripe AF order, short-range order, and block AF order make comparable contributions to the thermal conductance[47]. No magnetic quantum critical point, as in some superconducting or nonsuperconducting 2D iron pnictide systems, is observed in $BaFe_2S_{3-x}Se_x$[55-57]. In the S-rich samples with $x \leq 0.7$, a slight enhancement of the ordered moments with Se content may imply a strengthening of the electronic correlation.

In summary, we have studied the crystal structure, magnetic order, and electronic properties of the isovalent substitution series $BaFe_2S_{3-x}Se_x$ and constructed a comprehensive phase diagram in composition-temperature space. A first-order transition from the *Cmcm* structure with stripe AF order to the *Pnma* structure with block AF order has been identified for $x = 0.7 \sim 1.0$. For $x = 1$, short-range, 2D correlations of both stripe and block AF order are observed. The thermal activation gap extracted from resistivity measurements increases rapidly across the structural transition. Together, these results highlight the strong coupling between the structural, magnetic, and electronic degrees of freedom in $BaFe_2S_{3-x}Se_x$. We suggest that detailed studies of the electronic properties of intermediate compositions of $BaFe_2S_{3-x}Se_x$ under pressure may yield valuable new insights into the complex behavior in this system.

## Acknowledgements

Work at Sun Yat-Sen University was supported by the National Natural Science Foundation of China (Grant No. 11904414, 11904416), Natural Science Foundation of Guangdong (No. 2018A030313055), National Key Research and Development Program of China (No. 2019YFA0705700), the Fundamental Research Funds for the Central Universities (No. 18lgpy73), the Hundreds of Talents program of Sun Yat-Sen University, and the Pearl River Scholarship program. Work at UC Berkeley and Lawrence Berkeley Laboratory was supported by the Office of Science, Office of Basic Energy Sciences (BES), Materials Sciences and Engineering Division of the U.S. Department of Energy (DOE) under Contract No. DE-AC02-05-CH1231 within the Quantum Materials Program (KC2202) and BES. BAF acknowledges support from the College of Physical and Mathematical Sciences at Brigham Young University. We thank Gerald Morris, Bassam Hitti, and Donald Arseneau for valuable assistance during the μSR experiments. We thank Keith Taddei, Matt Tucker, and Michelle Everett for assistance with the neutron diffraction experiments at ORNL. The identification of any commercial product or trade name does not imply endorsement or recommendation by the authors.

*wangmeng5@mail.sysu.edu.cn
#benfrandsen@byu.edu

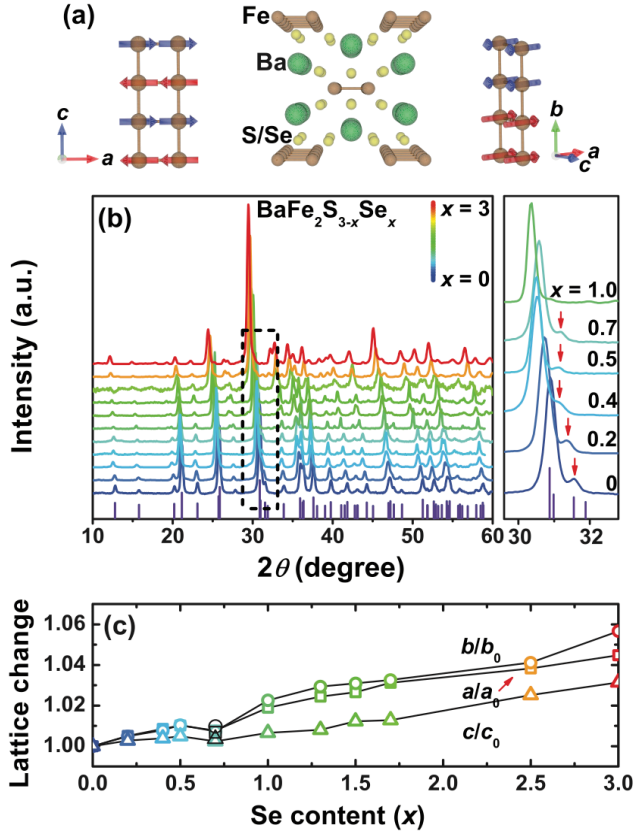

Fig. 1 (a) Crystal structure of BaFe$_2$S$_{3-x}$Se$_x$ and schematic views of Fe ladders with the coordinate systems and magnetic structures for $x = 0$ on left and $x = 3$ on right, respectively. We note that samples with $x \geq 1$ exhibit a distortion (not shown) in which the ladders tilt slightly out of the $b$-$c$ plane ($a$-$c$ plane in the coordinate system for $x = 0$). (b) X-ray diffraction patterns for BaFe$_2$S$_{3-x}$Se$_x$ ($x$ = 0, 0.2, 0.4, 0.5, 0.7, 1.0, 1.3, 1.5, 1.7, 2.5, 3.0), with the calculated peak positions for BaFe$_2$S$_3$ at the bottom and an expanded view in the dashed box to the right. (c) Lattice parameters extracted from the refinements, normalized by the values for BaFe$_2$S$_3$. The statistical errors are smaller than 1% and are covered by the symbols. The black symbols for $x = 0.7$ were obtained from a high-resolution x-ray diffractometer.

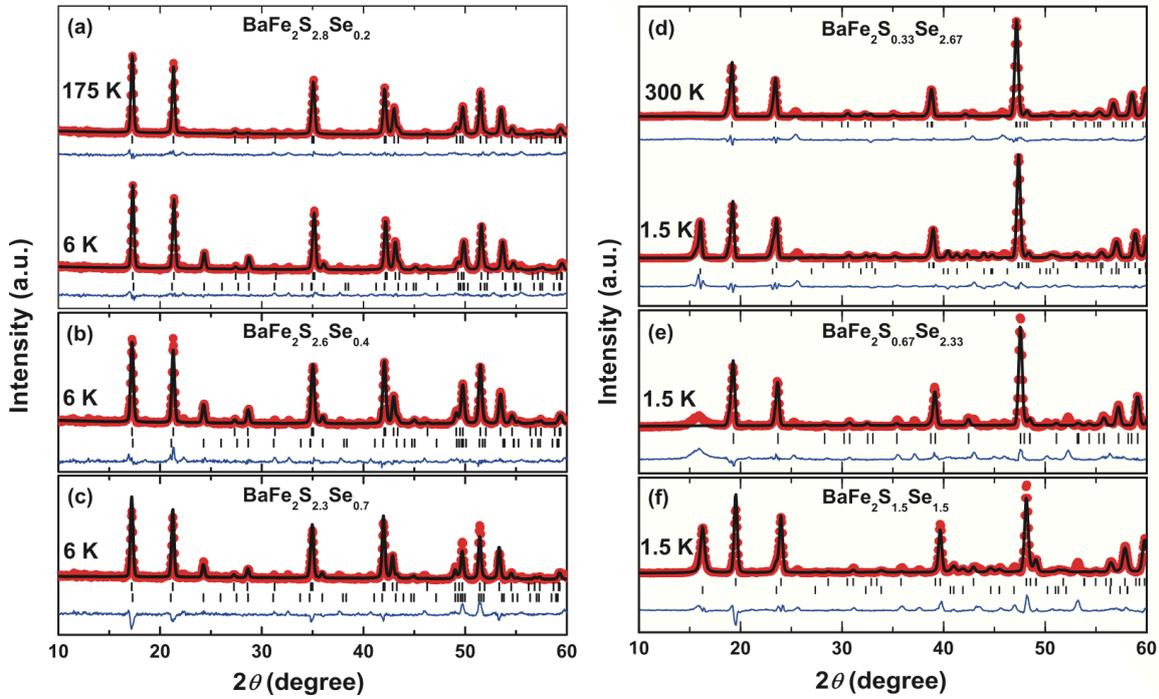

Fig. 2 (a) NPD patterns for BaFe$_2$S$_{2.8}$Se$_{0.2}$ at 175 and 6 K, (b) BaFe$_2$S$_{2.6}$Se$_{0.4}$ and (c) BaFe$_2$S$_{2.3}$Se$_{0.7}$ at 6 K, (d) BaFe$_2$S$_{0.33}$Se$_{2.67}$ at 300 and 1.5 K, (e) BaFe$_2$S$_{0.67}$Se$_{2.33}$ and (f) BaFe$_2$S$_{1.5}$Se$_{1.5}$ at 1.5 K. For each panel, the red dots represent the experimental data; the black line represents the calculated pattern; the black bars represent the calculated peak positions; and the blue line represents the difference between the observed and calculated patterns.

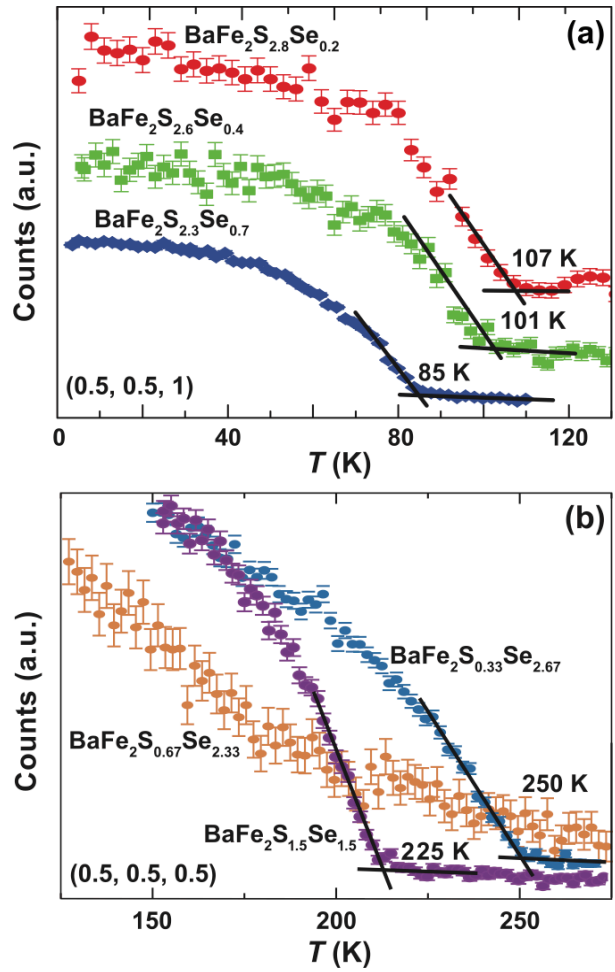

Fig. 3 (a) Order parameter measurements at $Q = (0.5, 0.5, 1)$ for the BaFe$_2$S$_{3-x}$Se$_x$ samples with $x = 0.2, 0.4, 0.7$. The data for BaFe$_2$S$_{2.3}$Se$_{0.7}$ with smaller error-bars are from single crystals. (b) Scattered intensity at $Q = (0.5, 0.5, 0.5)$ corresponding to the block AF order for $x = 1.5, 2.33$ and $2.67$. BaFe$_2$S$_{0.67}$Se$_{2.33}$ has a broad crossover to short-range block correlations.

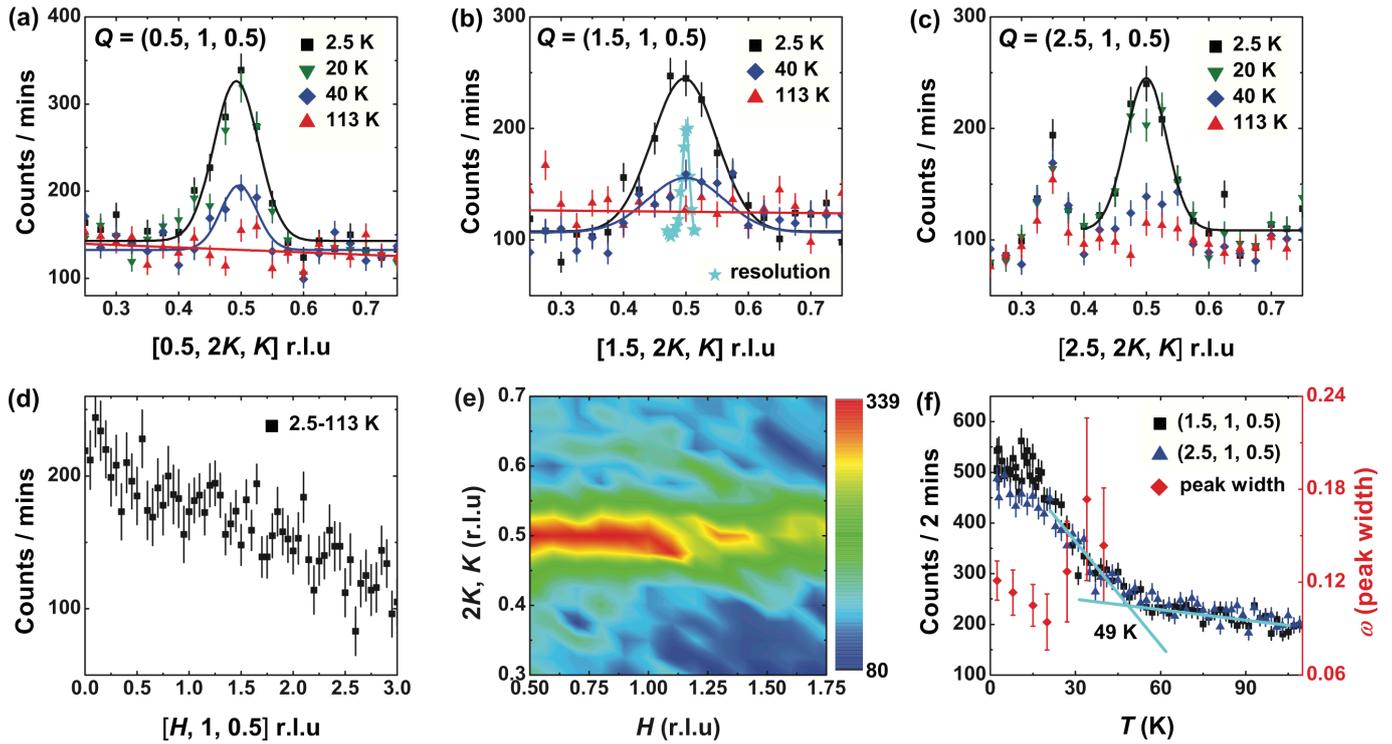

Fig. 4 Temperature dependence of the neutron diffraction measurements of single crystalline BaFe$_2$S$_2$Se. Magnetic peak scans at (a) $Q = (0.5, 1, 0.5)$, (b) $Q = (1.5, 1, 0.5)$, and (c) $Q = (2.5, 1, 0.5)$ in the *Pnma* coordinate system for different temperatures, where the solid lines are Gaussian or linear fits. The cyan stars in (b) represent a nuclear Bragg peak from higher-order wavelength ($\lambda/2$) neutrons. (d) Difference between the scans at 2.5 and 113 K along the [H, 1, 0.5] direction. (e) A map of the magnetic scattering intensity in the [H, 2K, K] plane at 2.5 K. (f) Magnetic order parameters at $Q = (1.5, 1, 0.5)$ and $Q = (2.5, 1, 0.5)$, where the solid lines are guides to the eye. The red diamond symbols represent the peak widths at $Q = (1.5, 1, 0.5)$ under different temperatures.

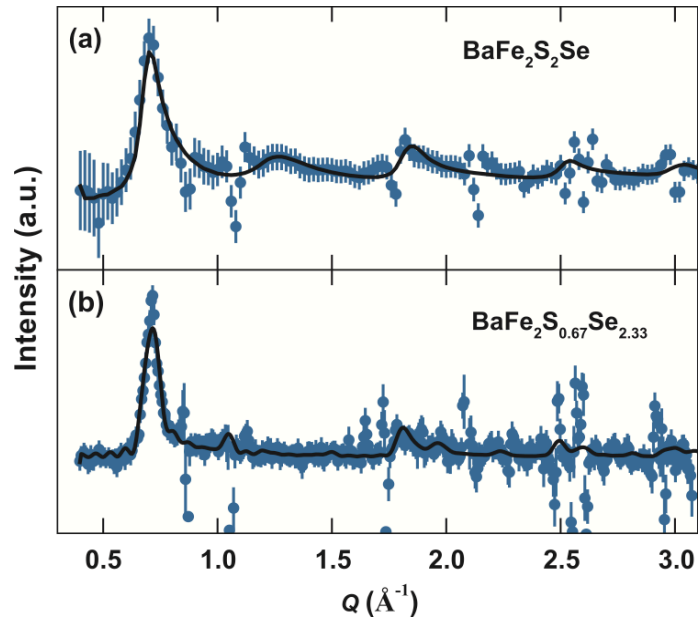

Fig. 5. Diffuse magnetic scattering for (a) BaFe$_2$S$_2$Se and (b) BaFe$_2$S$_{0.67}$Se$_{2.33}$. The data were obtained by subtracting high-temperature reference patterns (200 K for BaFe$_2$S$_2$Se and 300 K for BaFe$_2$S$_{0.67}$Se$_{2.33}$) from the low-temperature (2 K) diffraction patterns. The blue symbols are the data and the black curves are fits using various models of short-range correlations described in the main text.

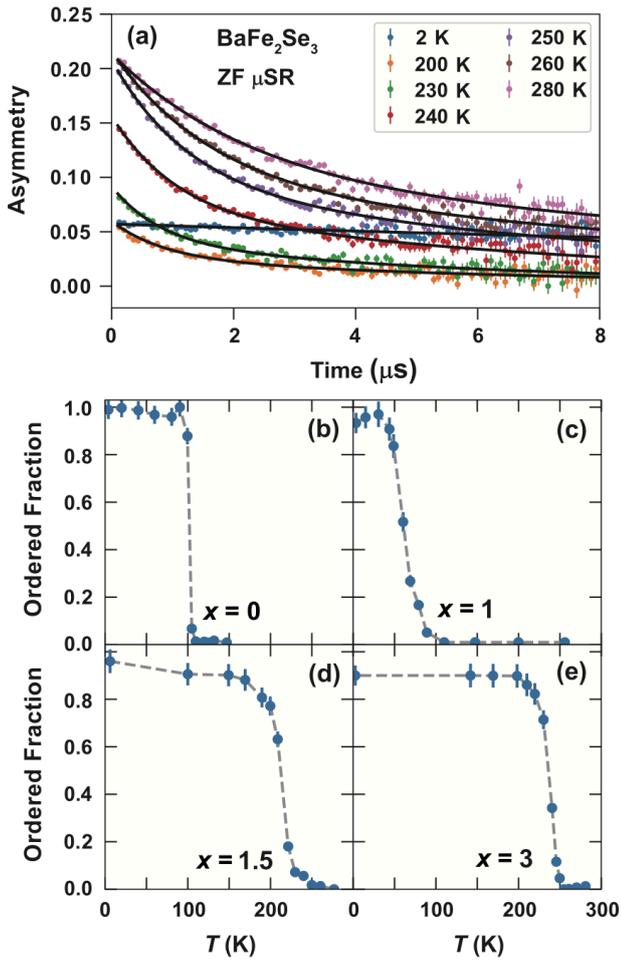
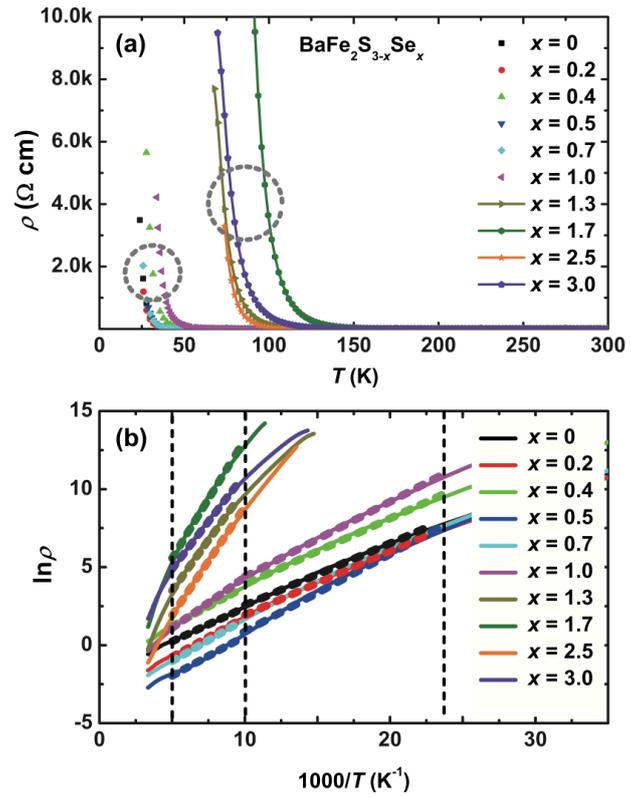

Fig. 6. (a) Representative zero-field μSR spectra for BaFe2S$_{3-x}$Se$_x$ showing the evolution of the asymmetry across $T_N$. (b-e) Magnetically ordered volume fraction for various compositions extracted from fits to the asymmetry spectra.

Fig. 7. (a) Resistivity versus temperature ($\rho - T$) curves for BaFe$_2$S$_{3-x}$Se$_x$, some of which have been included in a review paper[52]. (b) Fits to the $ln\rho$ vs. $1000/T$ curves using the activation-energy model. Solid (dashed) curves represent the data (fits), and the dashed vertical lines demarcate the fitting ranges.

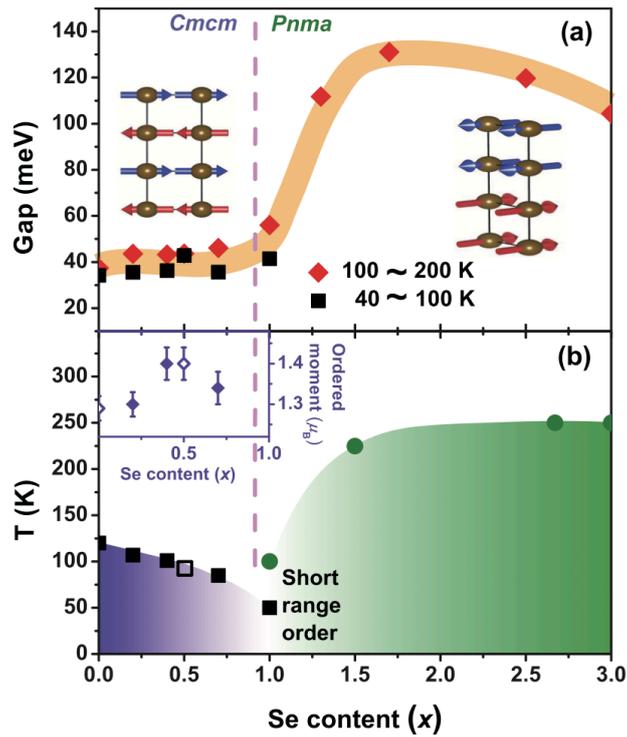

Fig. 8. Phase diagram for (a) the electronic state, where the errors (<1%) of the fitted thermal activation gaps are covered by the symbols; and (b) the magnetic state and ordered moment (inset), where the data for BaFe$_2$S$_3$ and BaFe$_2$S$_{2.5}$Se$_{0.5}$ come from our previous work [50,58].